\documentclass[journal,twoside,web]{ieeecolor}

\usepackage{tmi}
\usepackage{cite}
\usepackage{url}
\usepackage{amsmath,amssymb,amsfonts}
\usepackage{algorithmic}
\usepackage{graphicx}
\usepackage{textcomp}
\usepackage{multirow}
\usepackage{booktabs}
\usepackage{flushend} 

\DeclareMathOperator*{\argmin}{argmin}
\newcommand{\dg}{$^\circ$}

\begin{document}

\title{SE(3)-Equivariant and Noise-Invariant \\ 3D Rigid Motion Tracking in Brain MRI}

\author{Benjamin Billot, Neel Dey, Daniel Moyer, Malte Hoffmann, \\ Esra Abaci Turk, Borjan Gagoski, P. Ellen Grant, and Polina Golland
\thanks{B. Billot (corresponding author, e-mail: bbillot@mit.edu), N. Dey, and P. Golland are with MIT, Cambridge, USA.}
\thanks{D. Moyer is with Vanderbilt University, Nashville, USA.}
\thanks{M. Hoffmann is with Massachusetts General Hospital and Harvard Medical School, Boston, USA.}
\thanks{E. Abaci Turk, B. Gagoski, and P.E. Grant are with Boston Children’s Hospital and Harvard Medical School, Boston, USA.}
\thanks{This research is supported by NIH NIBIB 1R01EB032708, NIH NICHD R01HD100009, MIT CSAIL-Wistron Program, MIT-IBM Watson AI Lab, and MIT Jameel Clinic.}}

\maketitle


\begin{abstract}
Rigid motion tracking is paramount in many medical imaging applications where movements need to be detected, corrected, or accounted for. Modern strategies rely on convolutional neural networks (CNN) and pose this problem as rigid registration. Yet, CNNs do not exploit natural symmetries in this task, as they are equivariant to translations (their outputs shift with their inputs) but not to rotations. Here we propose EquiTrack, the first method that uses recent steerable SE(3)-equivariant CNNs (E-CNN) for motion tracking. While steerable E-CNNs can extract corresponding features across different poses, testing them on noisy medical images reveals that they do not have enough learning capacity to learn noise invariance. Thus, we introduce a hybrid architecture that pairs a denoiser with an E-CNN to decouple the processing of anatomically irrelevant intensity features from the extraction of equivariant spatial features. Rigid transforms are then estimated in closed-form. EquiTrack outperforms state-of-the-art learning and optimisation methods for motion tracking in adult brain MRI and fetal MRI time series. Our code is available at \url{https://github.com/BBillot/EquiTrack}.
\end{abstract}


\vspace{0.1cm}
\noindent \textbf{Published at IEEE transactions on Medical Imaging, 2024.}

\noindent \textbf{DOI: \url{10.1109/TMI.2024.3411989}}

\section{Introduction}
\label{sec:introduction}

\subsection{Motivation}

\IEEEPARstart{W}{ith} the advent of deep registration networks, alignment speed now approaches slice acquisition times in volumetric modalities like magnetic resonance imaging (MRI). Consequently, such registration methods unlock real-time prospective motion tracking, which has many biomedical applications. For example, prospectively detecting movements between consecutive slices in clinical brain MRI could help reducing the cost of motion artefacts, estimated to amount to 20\% of expenses for MRI exams\cite{andre_toward_2015}. Another compelling application is slice prescription in fetal MRI, where tracking uncontrollable movements could alleviate the need to repeatedly acquire the same scans until no motion occurs\cite{xu_fetal_2019}.

Accurately detecting 3D motion at high temporal resolution naturally leads to fast sequences, such as navigator scans\cite{white_promo_2010,malamateniou_motion-compensation_2013}. Due to time constraints, these images are acquired at low resolution and are often characterised by a low signal-to-noise (SNR) ratio. In this scenario, traditional tracking approaches based on the detection of predefined keypoints often fail since these might become challenging to identify \cite{bellekens_benchmark_2017}. Hence, it may be beneficial to \textit{learn} more robust features. Based on this idea, modern methods mostly rely on convolutional neural networks (CNNs) to directly infer pose changes across scans, effectively formulating motion tracking as rigid registration \cite{miao_cnn_2016,mohseni_salehi_real-time_2019}. 

While CNNs can perform accurate rigid registration, their design does not exploit natural symmetries in rigid motion: convolutional kernels are by construction equivariant to translations (i.e., their outputs shift with their inputs) but not to rotations. Thus, CNNs may extract different features for the same object in two poses, which is particularly misleading for motion tracking. Although equivariance can be learned with augmentation \cite{gruver_lie_2023} or semi-supervised learning\cite{chen_simple_2020}, these do not guarantee equivariance and are shown here to be suboptimal.

For this reason, there has been an increasing interest in building kernels that are equivariant to 3D rigid transforms, i.e., the SE(3) group. While initial methods worked for sets of discrete angles only \cite{cohen_group_2016,winkels_3d_2018,bekkers_roto-translation_2018}, full SE(3)-equivariance was introduced in \cite{weiler_3d_2018}, where convolution masks are expanded with learnable coefficients into bases of \textit{fixed} steerable filters (computed as spherical harmonics). Steerable equivariant CNNs (E-CNNs) are formed by stacking these filters into layers, separated with equivariant non-linearities \cite{weiler_3d_2018}. E-CNNs are trained by learning the linear expansion coefficients. 

In a preliminary work \cite{moyer_equivariant_2021}, we showed that E-CNNs obtain encouraging results for 3D motion tracking. However, experiments were only conducted on simulated data that ignored temporal intensity changes across scans. Here, testing E-CNNs on real time series shows that they do not have enough expressiveness to learn robustness against intensity changes. This is a major issue in medical imaging, where scans may differ due to scanner noise, motion artefacts, histogram shifting, etc.

\subsection{Contributions}
 
We present EquiTrack, an SE(3)-equivariant and intensity-invariant framework for 3D motion tracking. We leverage a hybrid architecture that decouples the processing of intensity and spatial features. This is achieved by using a denoising CNN to remove anatomically irrelevant intensity features from the noisy input data, and then employing an E-CNN to extract robust SE(3)-equivariant maps. Such representations (obtained in parallel for the input scans) are then collapsed onto point clouds, which by construction only differ by the unknown rigid transform. The latter is finally derived analytically with a differentiable closed-form algorithm.

We demonstrate EquiTrack on adult brain MRI and on two datasets of fetal brain MRI time series. Fetal time series are particularly challenging for motion tracking. First, rapid fetal movements necessitate fast low-resolution EPI sequences with low SNR and tissue contrast. Further, there are only a few identifiable spatial features that can be used to drive the registration. Finally, fetuses can be in arbitrary orientations.

Preliminary versions of EquiTrack were presented at MICCAI 2021 \cite{moyer_equivariant_2021} and at the short paper track of MIDL 2023 \cite{billot_equivariant_2023}. Here we extend the purely pose-equivariant method of \cite{moyer_equivariant_2021} into a complete framework for motion tracking in \textit{real} data by presenting a jointly noise-invariant and pose-equivariant architecture. Compared to \cite{billot_equivariant_2023}, we now evaluate EquiTrack against five additional baselines in three new experiments, including testing on adult and real fetal brain time series, and we report on extensive ablation and sensitivity studies.

Overall, EquiTrack outperforms state-of-the-art optimisation and learning-based rigid registration methods for 3D motion tracking. Importantly, our method produces fast motion estimates (under a second) and has the potential to be deployed in real-time clinical imaging pipelines.

\section{Related work}
\label{sec:related work}

\subsubsection{Optimisation-based motion tracking} 
Real-time motion tracking in volumetric modalities has a long history. Initial strategies have been applied to structural \cite{woods_automated_1998} and functional MRI \cite{cox_real-time_1999} by optimising a least-square image loss with gradient descent. Other solutions have proposed to detect motion by finding correlations between k-space signals collected in concentric strips around the origin \cite{pipe_motion_1999}. Subsequent strategies have advocated for the use of navigator scans to drive motion tracking with Kalman filters\cite{white_promo_2010}. All these methods have been developed for adult brain MRI, where head rotations rarely exceed $10^{\circ}$ \cite{pipe_motion_1999}, and often fail when applied to fetal imaging with much higher movement amplitudes\cite{malamateniou_motion-compensation_2013}. 

\subsubsection{Optimisation-based rigid registration}
Motion tracking can be formulated as a rigid registration problem between consecutive time frames of the same subject. Classical registration strategies rely on iterative optimisation of the minimisation of an image similarity loss at progressively finer resolutions~\cite{hill_medical_2001}. Most methods then vary with respect to the employed optimisation framework~\cite{jenkinson_global_2001} or similarity function~\cite{hajnal_detection_1995,wells_multi-modal_1996,roche_correlation_1998}. These approaches are widely implemented in rigid registration packages\cite{avants_reproducible_2011,modat_global_2014} but remain relatively slow.

\subsubsection{Landmark registration}
Another paradigm is based on detecting landmarks to estimate transforms (not necessarily rigid) with classical optimisation \cite{chui_new_2003,hill_medical_2001} or in closed-form\cite{arun_least-squares_1987,horn_closed-form_1987}. Such landmarks can be handcrafted\cite{tuytelaars_local_2008}, or rely on intensities\cite{montesinos_differential_1998} or geometric features\cite{pennec_framework_1997}, such as second order derivatives, which have been applied to fetal motion correction~\cite{scheinost_fetal_2018}. However, these strategies are sensitive to the quality of the extracted landmarks and thus of the data itself.

\subsubsection{Learning-based registration}
Modern rigid registration strategies mostly build on CNNs for fast and accurate alignments \cite{miao_cnn_2016}. In general, such methods directly regress the translation and rotation values with a CNN encoder followed by fully connected layers. While initial approaches were trained on ground truth transforms obtained by construction or from classical frameworks\cite{mohseni_salehi_real-time_2019,chee_airnet_2018}, a newer strategy inspired by optimisation frameworks directly minimises an unsupervised image similarity loss\cite{de_vos_deep_2019}. Reinforcement learning has also been used for rigid registration, but requires long processing times\cite{liao_artificial_2016}. More generally, it has been shown that learning methods do not generalise well to large movements \cite{wang_robust_2023}. Similar to our prior work \cite{moyer_equivariant_2021}, a recent method proposes to learn unsupervised keypoints from which rigid transforms are estimated in closed-form\cite{wang_robust_2023}. Effectively, \cite{wang_robust_2023} replaces the denoiser and E-CNNs of EquiTrack with a single CNN, a design shown here to be suboptimal for motion tracking.

\begin{figure*}[!th]
\centerline{\includegraphics[width=\textwidth]{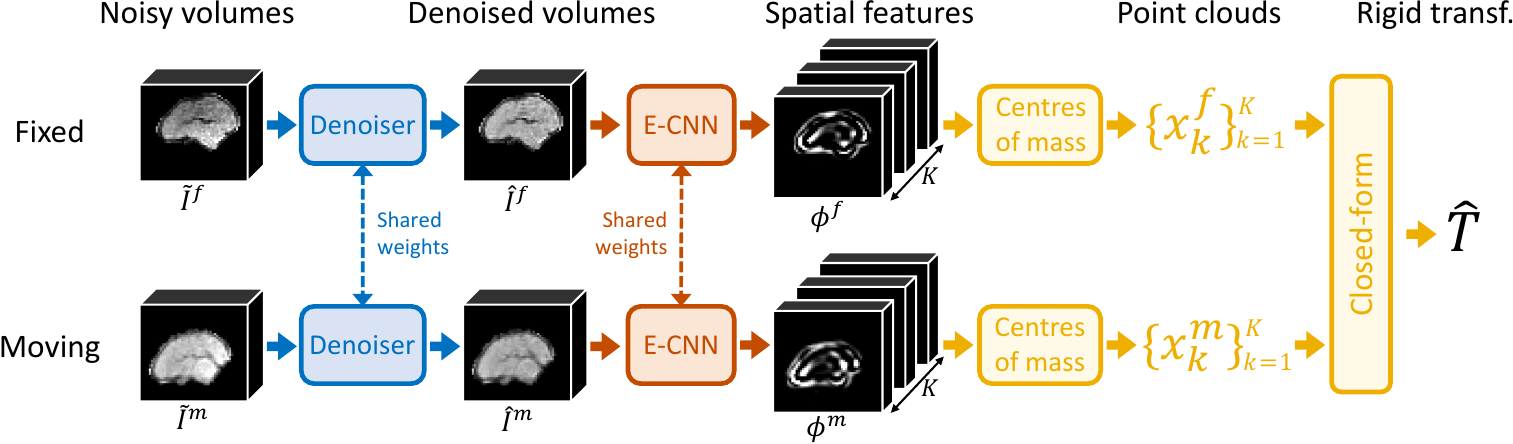}}
\caption{Overview of EquiTrack. The fixed and moving volumes are first processed with a denoising CNN that removes anatomically irrelevant intensity features (noise, histogram shifts, etc.), so that its outputs only differ by the unknown rigid transform. Crucially, we then use a steerable SE(3)-equivariant E-CNN to extract $K$ matching anatomical features across images. A rigid transform $\hat{T}$ is estimated by computing summary statistics (centres of mass), providing us with two corresponding point clouds that are registered with a differentiable closed-form algorithm\cite{horn_closed-form_1987}.}
\label{fig:overview}
\end{figure*}

\subsubsection{Equivariant networks}
Building equivariant CNNs has generated interest over the years\cite{freeman_design_1991,reisert_efficient_2008,duits_left-invariant_2011,janssen_hessian_2017}. These networks can be broadly classified along two axes: the type of objects they operate on, and the group of transformations they are equivariant to \cite{cohen_intertwiners_2018}. The first class of networks, known as group-CNNs, act on scalar fields. Equivariance is built by first applying rotated variants of the same filters (lifting convolutions), followed by several layers of group-convolutions and non-linearities \cite{cohen_group_2016}. While early approaches only tackled equivariance to $90^{\circ}$ rotations in 2D \cite{cohen_group_2016} and 3D \cite{winkels_3d_2018,worrall_cubenet_2018}, subsequent methods later used lifting convolutions with a finer angle discretisation \cite{bekkers_roto-translation_2018,lafarge_roto-translation_2021}. This strategy quickly becomes computationally prohibitive, as the number of convolutions increases linearly with the number of rotation angles. 

Meanwhile, \cite{worrall_harmonic_2017,cohen_steerable_2017} introduced steerable equivariant CNNs (E-CNN) for equivariance to \textit{continuous} rotations. In contrast to group-CNNs, such filters operate on vector fields. Equivariant convolutions were proven to be the most general equivariant linear maps between vector fields \cite{weiler_3d_2018}. Moreover, such convolutions can be expanded in a basis of steerable kernels, analytically derived by \cite{weiler_3d_2018} to be spherical harmonics modulated with radial functions\cite{weiler_3d_2018}. Although SE(3) is sufficient for rigid motion tracking, steerable E-CNNs have been recently extended to E(N) \cite{cesa_program_2022}.

E-CNNs have produced promising results in several medical imaging tasks including segmentation and anomaly detection in histopathology slices\cite{bekkers_roto-translation_2018, lafarge_roto-translation_2021}, pulmonary nodule detection in lung CTs \cite{winkels_3d_2018}, and fibre tract estimation in diffusion MRI \cite{bouza_higher_2021,elaldi_e3_2023}. To the best of our knowledge, this work is the first to apply E-CNNs to 3D motion tracking.

\subsubsection{Disentangling methods}
Disentangling methods separate intensity and anatomical features in parallel by mapping them to different sub-spaces \cite{huang_multimodal_2018,chartsias_disentangled_2019,liu_learning_2022}. Here, we process them successively to fully leverage the potential of E-CNNs.

\section{Methods}
\label{sec:methods}

\subsection{Problem formulation}

Our goal is to track the \textit{rigid} motion of an object (e.g., the fetal brain) using image time series. If the acquisition process is noise-free, there exists an unknown rigid transform $T$ such that the first 3D scan $I^f$ (\textit{fixed volume}) and any subsequent time frame $I^m$ (\textit{moving volume}) are related by:
\begin{equation}
I^m = T \circ I^f,
\label{eq:rigid}
\end{equation}
where $T\!\circ\!I^f$ is $I^f$ warped by $T$. Here we estimate $T$ by detecting matching features across volumes \cite{hill_medical_2001}. The feature extraction can be implemented as a network $\Phi$ that maps 3D images $I \in \mathcal{I}$ onto features maps $\{\phi_k\}_{k=1}^K$ \cite{moyer_equivariant_2021,wang_robust_2023}. For motion tracking, it is desirable for $\Phi$ to be \textit{SE(3)-equivariant}, so that features rotate and translate according to $T$:
\begin{equation}
\Phi(I^m) = \Phi(T \circ I^f) = T \circ \Phi(I^f).
\label{eq:equivariance}
\end{equation}
In other words, rigid-equivariance gives us fixed and moving features $\{\phi_k^f\}$ and $\{\phi_k^m\}$ that are in correspondence: $\phi_k^m = T \circ \phi_k^f$ for $k \in \{1,...,K\}$. Consequently, computing summary statistics of the two sets of features maps, for example by collapsing them onto their centres of mass, yields two point clouds $\{x_k^f\}$ and $\{x_k^m\}$ that are related by $x_k^m = Tx_k^f$ for $k \in \{1,...,K\}$. Thus, we can estimate $T^{-1}$ by deriving the rigid transform that optimally projects $\{x_k^m\}$ to $\{x_k^f\}$. This is a well-studied problem that is solved in closed-form with a differentiable expression \cite{arun_least-squares_1987,horn_closed-form_1987}.

However, the initial assumption of a noise-free acquisition process does not hold in practice, where the fixed and moving scans often present different intensity distributions (noise, artefacts, histogram shifts, bias field, etc.). We will see in Section~\ref{sec:experiments} that in practice steerable E-CNNs are not expressive enough to learn real-world intensity invariance. Thus, $\{\phi_k^f\}$ and $\{\phi_k^m\}$ estimated from images are not necessarily in correspondence, leading to poorer estimates of $T$ as intensity changes grow larger. Meanwhile, regular CNNs have been shown to learn invariance to intensity changes given appropriate training data\cite{ilesanmi_methods_2021,billot_synthseg_2023,hoffmann_synthmorph_2022,hoffmann_anatomy-specific_2023}. Hence, we employ a denoising CNN $\Psi$ to map the noisy inputs $\Tilde{I}^m$ and $\Tilde{I}^f$ onto a common noise-free intensity space: $I^f = \Psi(\Tilde{I}^f)$ and $I^m = \Psi(\Tilde{I}^m)$. This strategy decouples the processing of intensity features from the motion tracking procedure, which can be applied as described above.

The proposed method is illustrated in Fig.\ref{fig:overview}. The rest of section \ref{sec:methods} presents each module and the training procedure.

\subsection{Denoising CNN}
\label{sec:denoiser}

The first step of EquiTrack consists of removing the anatomically irrelevant features from the noisy input images, such that the outputs of $\Psi$ only differ by the pose of the represented object. Let $\zeta : \mathcal{I} \rightarrow \Tilde{\mathcal{I}}$ be an operator modelling the noise process happening during image acquisition, i.e., $\Tilde{I}^f = \zeta(I^f)$ and $\Tilde{I}^m = \zeta(I^m)$. The noise operator $\zeta$ is stochastic: applying it twice separately on the same input yields two different outputs. The purpose of the denoiser $\Psi$ can then be formulated as learning to undo $\zeta$: $\Psi\circ\zeta=Id$, such that $\Psi(\Tilde{I}^f)$ and $\Psi(\Tilde{I}^m$) (i.e., the inputs of $\Phi$) respect the assumption in \eqref{eq:rigid}:
\begin{equation}
\Psi(\zeta(I^m)) = I^m = T \circ I^f = T \circ \Psi(\zeta(I^f)).
\label{eq:invariance}
\end{equation}

In practice, $\Psi$ is not perfect and there will be residual noise remaining in its outputs: $\Psi(\Tilde{I}) = \Psi(\zeta(I)) = \hat{I} \approx I$. As a result, $T \circ \Psi(\Tilde{I}^f) = T \circ \hat{I}^f$ now becomes an approximation of $\Psi(\Tilde{I}^m) = \hat{I}^m$. Nevertheless, we find that $T \circ \hat{I}^f$ is a good approximation of $\hat{I}^m$. Moreover, the residual differences will be tackled later by using a robust estimation method when deriving the rigid transform $T$.

More generally, we emphasise that undoing $\zeta$ is a crucial step. Indeed, $\Psi$ enables us to learn approximate noise invariance, which is extremely difficult to achieve for $\Phi$ since, as we will show, its limited number of learnable parameters does not provide it with enough learning capacity. Overall, the denoising CNN can be seen as an efficient way to decouple the processing of intensity artefacts from anatomical features to fully leverage the potential of the steerable E-CNN.

\subsection{SE(3)-equivariant feature extraction}

We now seek to efficiently extract SE(3)-equivariant features from denoised images in a differentiable manner, leading us to use steerable SE(3)-equivariant CNNs. We now describe the construction of $\Phi$, following the presentation of \cite{weiler_3d_2018}.

\subsubsection{Non-scalar fields and representations} 
Regular CNNs generally produce scalar-valued outputs from scalar fields. Yet, more expressive features can be obtained from higher-order fields \cite{cohen_steerable_2017}. A scalar field $\phi:\mathbb{R}^3\rightarrow\mathbb{R}$ transforms simply with rotation $r$ and translation $t$: the output at location~$x$ is given by $\phi(r^{-1}(x-t))$. However, if $\phi$ is a higher-order field (e.g., a vector field $\mathbb{R}^3\rightarrow\mathbb{R}^3$), it transforms as $\rho(r)\phi(r^{-1}(x-t))$, where $\rho(r)$ is a SO(3) \textit{representation} depending on the field order and describing how the dependant channels of $\phi$ transform with $r$. More generally, let $\phi: \mathbb{R}^3\rightarrow\mathbb{R}^K$ be a mix of several subfields with different orders. It can be shown that for any SO(3) representation, there exists a basis that yields a block-diagonal matrix representation:
\begin{equation}
\rho(r) = Q^{-1} \left[ \bigoplus {}_{l=0}^L D_l(r) \right] Q,
\label{eq:irrep}
\end{equation}
where $Q$ is the change of basis, $D_l$ is the \textit{irreducible} SO(3) representation for the $l$-subfield with size $\mathbb{R}^{(2l+1) \times (2l+1)}$, $L$ is the highest subfield order, and $\oplus$ constructs a block-diagonal matrix with $D_l(r)$ as blocks. Intuitively, $Q$ lets $\rho$ live in a basis where each subfield transforms independently for rotations.

\subsubsection{Equivariant kernels} 
Let $\phi_n$ and $\phi_{n+1}$ be two fields before and after the $n$-th layer of a network, with arbitrary representations $\rho_n$ and $\rho_{n+1}$. A linear mapping $\kappa: \mathbb{R}^3 \rightarrow \mathbb{R}^{K_n \times K_{n+1}}$ between the two fields is SE(3)-equivariant if and only if it is a convolution using rotation-steerable kernels \cite{weiler_3d_2018}:
\begin{equation}
\kappa(rx) = \rho_{n+1}(r)\kappa(x)\rho_n(r)^{-1}.
\label{eq:steerable}
\end{equation}
This linear constraint implies that SE(3)-equivariant kernels form a vector space, for which we can find a basis.

\subsubsection{Building a basis of equivariant kernels}
Assuming that $\phi_n$ has been made irreducible by sorting its channels, (i.e., $\rho_n$ is block-diagonal), $\kappa$ can be decomposed into sub-kernels $\kappa_{lj}$ between the $l$-th order subfield of $\phi_n$ and $j$-th order subfield of $\phi_{n+1}$. Applying \eqref{eq:steerable} to each kernel block yields:
\begin{equation}
\kappa_{lj}(rx) = D_j(r)\kappa_{lj}(x)D_l(r)^{-1},
\label{eq:block-kernels}
\end{equation}
where $\kappa^{lj}:\mathbb{R}^3\rightarrow\mathbb{R}^{(2l+1) \times (2j+1)}$. After vectorising (i.e., flattening), \eqref{eq:block-kernels} becomes $\text{vec}(\kappa_{lj}(rx)) = \left[D_j \otimes D_l \right] \text{vec}(\kappa_{lj}(x))$, where $\otimes$ is the tensor product. Note that $\left[D_j \otimes D_l \right]$ is a representation that factors as in \eqref{eq:irrep}, with irreducible representations of orders $|j-l| \leq J \leq j+l$ \cite{cohen_steerable_2017}. Thus after vectorising and changing basis with $\eta_{lj}(rx) = Q \; \text{vec}(\kappa_{lj}(rx))$, we obtain:
\begin{equation}
\eta_{lj}(rx) = \left[ \bigoplus {}_{J=|j-l|}^{j+l} D^J(r) \right] \eta_{lj}(x).
\label{eq:sh_constraint}
\end{equation}
The unique solution for \eqref{eq:sh_constraint} are spherical harmonics $S^J(x) = (S^J_{-J}(x), ..., S^J_J(x))$ that are modulated with arbitrary radial functions $\{\varphi_m\}$ to form a basis of equivariant kernels \cite{weiler_3d_2018}:
\begin{equation}
\! \! \! \! \eta_{lj,J,m}(x) \! = \! \varphi_m(\|x\|)\,S^J\! \! \left(\!\frac{x}{\|x\|}\! \right)\!, \; \; \varphi_m(\|x\|) \! = \! e^{\frac{-(\|x\| - m)^2}{2\sigma^2}}\! \! . \;
\label{eq:sh_solution}
\end{equation}
After unvectorising the $\eta_{lj,J,m}$ and mapping back to the original basis, the final kernels are obtained as linear combination $\kappa_{lj}= \sum_{J,m}\nu_{lj,J,m}\,\kappa_{lj,J,m}$ with learnable $\nu_{lj,J,m}$ coefficients.

\subsubsection{Steerable E-CNN} 
Steerable E-CNNs are formed by interleaving layers of the above-described filter banks with equivariant non-linearities, implemented as multiplicative equivariant scalar fields \cite{weiler_3d_2018}. While steerable kernels have been derived in the continuous case, they can be easily discretised by sampling on $\mathbb{Z}^3$, followed by low-pass filtering with Gaussian smoothing to avoid aliasing \cite{weiler_3d_2018}. During training, all filters $\kappa_{lj,J,m}$ remain fixed and we only learn the coefficients $\nu_{lj,J,m}$. Here, we implement $\Phi$ as a steerable E-CNN with scalar inputs (the images) and outputs (the $K$ feature maps).

\subsection{Closed-form derivation of T}
\label{sec:svd}

After processing $\Psi(\Tilde{I}^f)$ and $\Psi(\Tilde{I}^m$) with the SE(3)-equivariant steerable E-CNN $\Phi$, we obtain two sets of features $\{\phi_k^f\}$ and $\{\phi_k^m\}$, where ideally each channel pair is related by $T$: $\phi_k^m = T \circ \phi_k^f, k \in \{1, ..., K\}$. We now seek to estimate $T$. 

First, we compute summary statistics of all feature maps~$\{\phi_k\}$ by finding their centre of mass $\{x_k\}$, given by:
\begin{equation}
x_k = \frac{1}{|\Omega|} \sum_{x \in \Omega} x \phi_k(x) \quad k \in \{1,...,K\},
\label{eq:CoM}
\end{equation}
where $\Omega \subset \mathbb{Z}^3$ is the compact spatial support of $\phi_k$, and $|\Omega|$ is its cardinality. This provides us with two corresponding point clouds: $x_k^m = T x_k^f$, $k \in \{1, ..., K\}$, where $T$ can be easily estimated by solving this over-determined linear system. 

However, this system is only an approximation due to discretisation errors and residual intensity differences between the denoised volumes, which propagate through the extracted features and point clouds. Thus, to gain robustness to slight misalignments, we formulate the point cloud regression as a least squares problem:
\begin{equation}
\hat{T} = \argmin_T \sum_k \|x_k^m - Tx_k^f \|^2.
\label{eq:SSD}
\end{equation}
Multiple closed-form solutions have been proposed for \eqref{eq:SSD}\cite{kabsch_solution_1976,arun_least-squares_1987,horn_closed-form_1987}. While \eqref{eq:SSD} considers all $K$ extracted points equally, 
these points might not be predicted by the E-CNN with the same confidence. Hence, we introduce channel weights $\{w_k\}_{k=1}^K$ to account for the reliability of the corresponding points, and the previous equation now becomes:
\begin{equation}
\hat{T} = \argmin_T \sum_k w_k \|x_k^m - Tx_k^f \|^2.
\label{eq:weighted_SSD}
\end{equation}
Here, we use the magnitude of the activations of $\Phi$ as a proxy for confidence: high activations indicate strong match between the extracted features and the representations learned by the network, while low activations denote weaker correlations that should be regarded as less reliable. Specifically, we define two sets of image-specific weights $\{w^f_k\}$ and $\{w^m_k\}$ as the sum of absolute activations $\Tilde{w}^f_k = \sum_\Omega |\phi_k^f|$ normalised across channels $w^f_k = \Tilde{w}^f_k / \sum_k \Tilde{w}^f_k$ (similarly for $\{w^m_k\}$). We then obtain the final channel weights $\{w_k\}$ by combining the image-specific weights $w_k = \! w_k^f \! \times \! w_k^m$.

More generally, Eq. \eqref{eq:weighted_SSD} also has a closed-form solution \cite{horn_closed-form_1987}. First, the optimal translation $\hat{t}$ is estimated by subtracting the weighted centroids of the moving and fixed point clouds: \mbox{$\hat{t} = \Bar{x}^m - \Bar{x}^f$} with \mbox{$\Bar{x}^f = \sum_k w_kx_k^f$} and \mbox{$\Bar{x}^m = \sum_k w_kx_k^m$}. The rotation $\hat{R}$ is then derived as follows. We recenter all points around their centroid \mbox{$\Bar{x}^f_k = x^f_k - \Bar{x}^f$} and stack them into matrices $\Bar{X}^f$ and $\Bar{X}^m$. If \mbox{$W = \text{diag}(w_1,...,w_k)$}, then \mbox{$\hat{R} = VU^T$}, where $U\Lambda V^T$ is the singular value decomposition (SVD) of the weighted cross-correlation matrix \mbox{$\Sigma = (\Bar{X}^m)^TW\Bar{X}^f$}.

\subsection{Learning and training losses}

EquiTrack is fully differentiable, yet end-to-end training suffers from vanishing gradients through $\Phi$ (due to the gated nonlinearities \cite{weiler_3d_2018}), making the training of $\Psi$ unstable. Thus, we train $\Psi$ and $\Phi$ separately and combine them at inference.

The denoiser $\Psi$ is trained to undo the effects of a given operator $\zeta$ (described later in this section) by minimising the L2 loss: $\mathcal{L}_\Psi = \sum_{x\in\Omega} \| I(x) - \Psi(\zeta(I(x)))\|^2$.

The steerable E-CNN $\Phi$ is trained on simulated pairs obtained by rigidly deforming an anchor volume twice with $T_1$ and $T_2$. Here, we employ an unsupervised image loss: $\mathcal{L}_\Phi =\sum_{x\in\Omega} \| T_2 \circ I(x) - \hat{T} \circ (T_1 \circ I(x)) \|^2$. While we could also use a supervised loss (since the ground truth transforms are known), we show that $\mathcal{L}_\Phi$ yields slightly better results.

\subsection{Implementation details}

\subsubsection{Noise process $\zeta$}
\label{sec:zeta}
For our application, we model $\zeta$ as an MRI-specific data augmentation process. Following \cite{billot_synthseg_2023}, we include corruption with a bias field $B$, histogram shifting with a $\gamma$-exponentiation, and injection of scanner noise $\xi$, which can all widely vary across scans, especially in fetal imaging:
\begin{equation}
\zeta(I, B, \gamma, \xi) = (I \times B)^{\gamma} + \xi,
\label{eq:augmentation}
\end{equation}
where $\gamma$ is a scalar and intensities are assumed to be in [0, 1], so that the $\gamma$-exponentiation leaves them in [0, 1]. Briefly, $B$ represents spatially varying inhomogeneities in the magnetic field of MRI scanners \cite{ashburner_unified_2005}, and is obtained by sampling a small grid (here of size $4\times4\times4$) from $\mathcal{N}(0, \sigma_B^2)$ and linearly upsampling it to the image size. Scalar $\gamma$ is drawn from $\mathcal{N}(0, \sigma_\gamma)$, and the noise field $\xi$ is sampled from $\mathcal{N}(0, \sigma_\xi^2)$ at every voxel. To increase variability, $\sigma_B$ and $\sigma_\xi$ are drawn from uniform distributions $\mathcal{U}[0, {\sigma_B}_{\max}]$ and $\mathcal{U}[0, {\sigma_\xi}_{\max}]$. 

\subsubsection{Architecture and training parameters}
\label{sec:architecture}

We implement the denoiser $\Psi$ as a 3D UNet \cite{ronneberger_u-net_2015} with 4 levels, each with 2 convolution layers of 32 $3\!\times\!3\!\times\!3$ kernels, ReLU activations, and batch normalisation \cite{ioffe_batch_2015}. Inspired by \cite{kamath_we_2023,billot_robust_2023}, we remove the top skip-connection, which affected reconstructions by reintroducing early noisy features at the last stage of the UNet.

The pose regressor $\Phi$ is as a steerable E-CNN with 5 SE(3)-equivariant layers using $5\!\times\!5\!\times\!5$ kernels (i.e., $\varphi_m(x>5) = 0$), separated by equivariant ReLU non-linearities \cite{weiler_3d_2018}. The inputs are the scalar-valued images, and we set the outputs to be 64 scalar-valued channels. Finally, the 4 intermediate layers use 4, 16, and 16 fields of orders 0, 1, and 2, respectively.

These architectures were optimised on a validation set. The networks $\Psi$ and $\Phi$ are trained separately with the Adam optimiser \cite{kingma_adam_2014} (learning rate of $10^{-5}$) for 50,000 and 10,000 iterations, respectively. Models are then selected based on validation scores, and aggregated into one framework for fast inference on a GPU (Nvidia Titan XP) using PyTorch \cite{paszke_pytorch_2019}.

\subsubsection{Augmentation parameters}

The inputs of $\Phi$ are obtained by sampling rigid transforms from $[-180^\circ, 180^\circ]$ and $[-20, 20]$ voxels for rotations and translations. The parameters of $\zeta$ are set to ${\sigma_B}_{\max}= 0.3$, $\sigma_\gamma = 0.2$, ${\sigma_\xi}_{\max} = 0.05$. These values can sometimes lead to slightly unrealistic augmentations, which has been shown before to increase robustness \cite{billot_synthseg_2023}. Further, we augment the inputs of $\Psi$ with spatial transforms (the same as for $\Phi$), as well as also apply $\zeta$ to the inputs of $\Phi$ in an attempt to make it learn noise invariance ($\mathcal{L}_\Phi$ is computed on the \textit{clean} images). Augmentations are applied online \cite{perez-garcia_torchio_2021} to increase the variability of the training data, and thus the robustness of the resulting network.

\section{Experimental Setup}

\subsection{Datasets}
\label{sec:datasets}

\subsubsection{Adult} We first evaluate EquiTrack on adult brain MRIs from the Human Connectome Project \cite{van_essen_human_2012}. We use 100 subjects to roughly match the size of the subsequent fetal datasets (with fewer subjects but more scans). Volumes are skull-stripped T1-weighted sequences at 0.7mm isotropic resolution that we resample to 1.5mm resolution. Preprocessing includes padding to a $128^3$ size and rescaling of intensities to [0,1] with robust min-max normalisation (using $1^\text{st}$ and $99^\text{th}$ percentiles). We use splits of 60/10/30 for training/validation/testing.

\subsubsection{Fetal-I} We use a dataset of 55 whole-uterus 3D MRI time series with fetuses of gestational age from 25 to 35 weeks. Mothers are in a supine or lateral position. Scans are acquired on a 3T Skyra Siemens scanner using multi-slice single-shot gradient echo EPI sequences at 3mm isotropic resolution (TR = [5-8]ms, TE = [32-38]ms, $\alpha$ = 90\dg). The mean image grid size is $120\!\times\!120\!\times\!80$, which we crop and pad to $96^3$. Intensities are rescaled to [0,1]. Automatic brain masking is applied and masks are dilated by 4 voxels to account for segmentation uncertainty. We split the time series into 30 for training (N=764 scans in total), 5 for validation (N=152), and 20 for testing (N=633).

\subsubsection{Fetal-II} We include a dataset of 23 \textit{clinical} fetal MRI time series from 14 subjects. Importantly, all subjects are Chiari-II fetuses, characterised by cerebral and spinal malformations \cite{kuhn_chiari_2023}. Further, these acquisitions use a gradient echo EPI sequence that is different from Fetal-I (TR=6.5s, TE=37ms, 2mm isotropic resolution). Volumes are preprocessed as before (see Appendix~\ref{sec:app_segm} for more details on the skull stripping step). These scans (N=614) are only used for \textit{testing}, in order to evaluate the generalisation of our method.

\begin{figure*}[th!]
\centerline{\includegraphics[width=\textwidth]{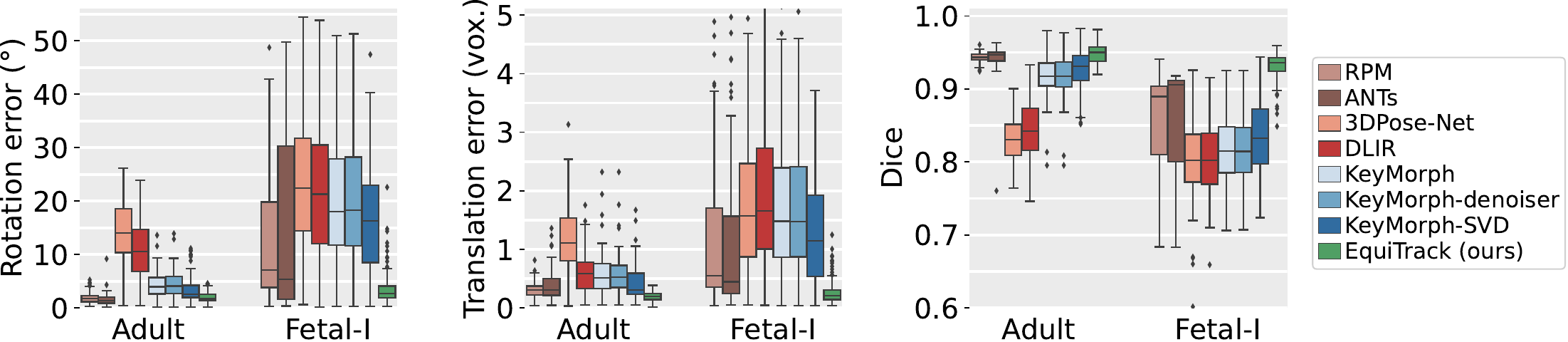}}
\caption{Box plots for rotation errors, translation errors, and Dice scores on simulated pairs from the Adult and Fetal-I datasets. EquiTrack is significantly better than all other methods at the 5\% level (Bonferroni-corrected two-sided non-parametric Wilcoxon signed-rank test), except for RPM and ANTs in terms of rotation error and Dice in adults (no statistically significant difference).}
\label{fig:boxplots}
\end{figure*}

\begin{figure*}[!t]
\centerline{\includegraphics[width=\textwidth]{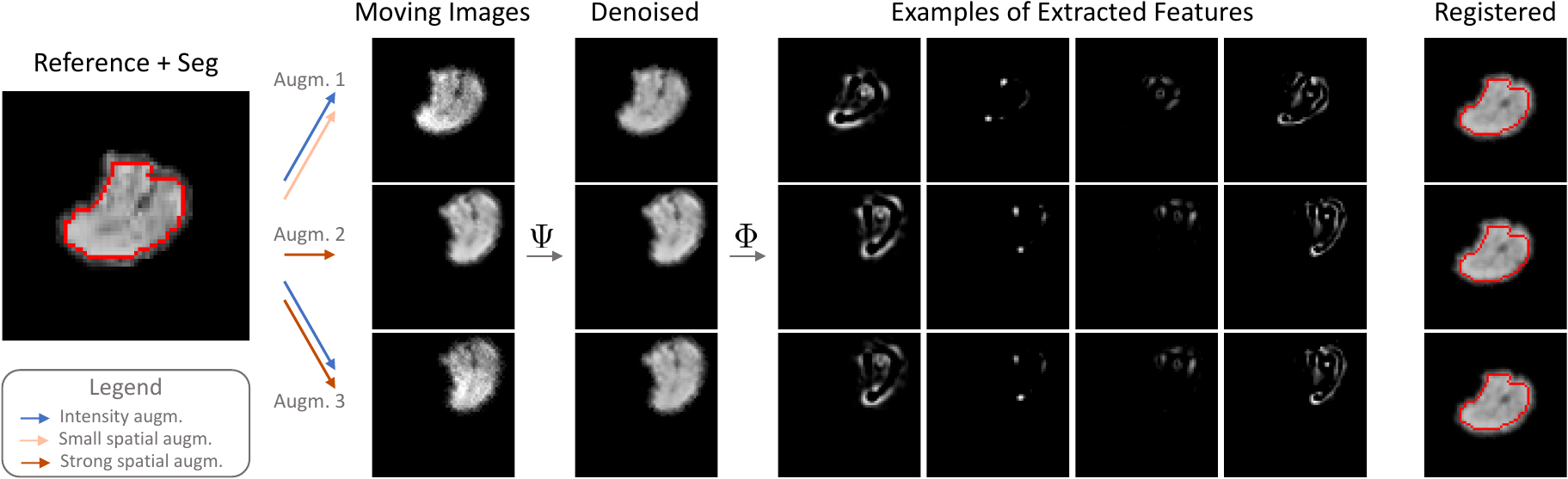}}
\caption{Example of the intermediate representations of EquiTrack. A reference scan (left) is either augmented with a large intensity transform (top row), a large spatial deformation (middle), or both (bottom). The denoiser $\Psi$ then removes noisy intensity features when appropriate (note how the middle example is left intact as no noise was added to it). After having removed intensity discrepancies across scans, the E-CNN $\Phi$ can now process volumes that only differ by their pose, which ensures the extraction of matching features across poses due to the SE(3)-equivariance of $\Phi$. These features are then used to accurately estimate rigid transforms back to the original volume whose brain mask is outlined in red for reference.}
\label{fig:equivariance}
\end{figure*}

\subsection{Baseline methods}

We test seven variants of five competing baselines. We train each learning method twice, and select the model with the best validation scores.

\subsubsection{Robust Point Matching (RPM) \cite{jian_robust_2011}} This rigid registration technique extracts surface meshes (e.g., with the marching cubes algorithm~\cite{lorensen_marching_1987}) from segmentations of the objects to align. These meshes are then converted into continuous distributions using Gaussian Mixture Models, which are iteratively aligned at different scales with numerical optimisation (using a quasi-Newton algorithm). We select the parameters that maximise the validation scores: stride of 2 for marching cubes, and 3 levels of optimisation (scale parameter = 0.5, 0.2, 0.1).

\subsubsection{ANTs \cite{avants_reproducible_2011}} The Advanced Normalizing Tools (ANTs) package is the state-of-the-art in optimisation-based medical image registration. We tune ANTs extensively on the validation set, which strongly improves results over the default hyperparameters. Briefly, we use a rigid 8-stage scheme, where the first seven steps minimise the L2 image loss (100 iterations, learning rate of 0.2), and the last stage uses $5\!\times\!5\!\times\!5$ patch-wise normalised cross correlation (10 iterations, learning rate of 0.05). The first four steps are at half resolution with Gaussian blurring of $\sigma\!=\!3,2,1,0$, respectively. We also add blurring at full resolution for levels 5 and 6 ($\sigma=2,1$, respectively).

\subsubsection{3DPose-Net \cite{mohseni_salehi_real-time_2019}} This method rigidly registers scans by regressing the six transform parameters (one translation and rotation for each axis) using three convolution layers followed by a regression head of three fully connected layers. Volumes are concatenated before being fed to the network, which is trained with the same L2 image loss as EquiTrack (slightly better results than the original geodesic loss).

\subsubsection{DLIR \cite{de_vos_deep_2019}} This strategy is similar to 3DPose-Net except that input scans are processed in parallel by the CNN (like in EquiTrack) and the extracted features are concatenated before the regression head. For a fair comparison, we match the number of layers to that of 3DPose-Net, which slightly improves the performance of the original implementation.

\subsubsection{KeyMorph \cite{wang_robust_2023}} KeyMorph is a state-of-the-art learning method for rigid registration. It learns to extract keypoints in an unsupervised fashion using the L2 image loss and aligns them with a closed-form algorithm. KeyMorph differs from EquiTrack in three aspects: a single CNN replaces the denoising CNN and E-CNN; this CNN is pre-trained with self-supervised learning; and uses a different closed-form solution to estimate transforms, derived by taking the derivative of \eqref{eq:SSD}. We implement KeyMorph and a variant that uses our closed-form algorithm (KeyMorph-SVD). We also test another variant to study the effect of preprocessing KeyMorph's inputs with the trained denoiser of EquiTrack (KeyMorph-denoiser).

\begin{figure*}[!t]
\centerline{\includegraphics[width=\textwidth]{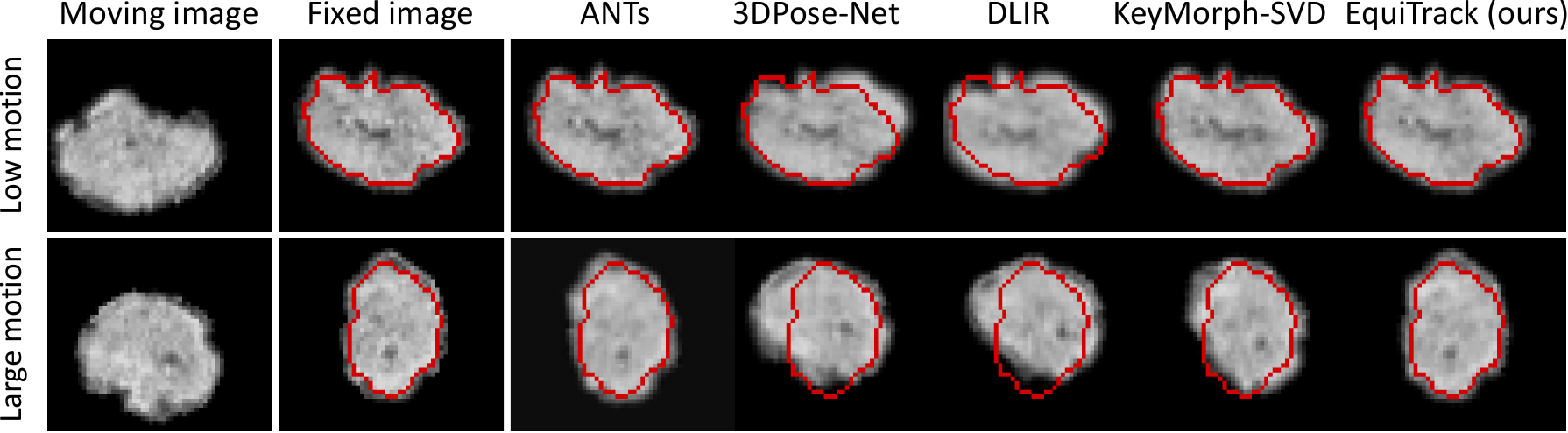}}
\caption{Sample registrations for representative methods on 3D pairs simulated from Fetal-I with small and large movements. The brain mask of the fixed image is shown in red for reference. Smoothness in registered images is due to interpolation. For small movements (top), all methods yield accurate or adequate results. In the case of large motion (bottom), only EquiTrack and ANTs (although less accurate) produce correct alignments.}
\label{fig:example_exp1}
\end{figure*}

\begin{table*}[t]
\centering
\caption{Number of parameters for all methods (in order of magnitude), and training/inference times on Fetal-I. \\ The denoiser and E-CNN components of EquiTrack are trained separately.}
\setlength{\tabcolsep}{7.5pt}
\begin{tabular}{|l | c | c | c | c | c | c c c|}
\hline
\multirow{2}{*}{} & \multirow{2}{*}{RPM} & \multirow{2}{*}{ANTs} & \multirow{2}{*}{3DPose-Net} & \multirow{2}{*}{DLIR} & \multirow{2}{*}{KeyMorph} & \multicolumn{3}{c|}{EquiTrack} \\
& & & & & & Denoiser alone & E-CNN alone & Combined \\
\hline
Number of total parameters & - & - & 2$\times$10$^8$ & 1$\times$10$^8$ & 6$\times$10$^6$ & 5$\times$10$^6$ & 2$\times$10$^6$ & 7$\times$10$^6$ \\ 
Number of learnable parameters & - & - & 2$\times$10$^8$ & 1$\times$10$^8$ & 6$\times$10$^6$ & 5$\times$10$^6$ & 8$\times$10$^4$ & 5$\times$10$^6$ \\ 
Time per training iteration (s) & - & - & 0.30 & 0.22 & 0.59 & 0.32 & 0.52 & - \\
Inference time (s) & 4.2 & 7.3 & 0.15 & 0.12 & 0.37 & 0.17 & 0.34 & 0.51 \\ 
\hline
\end{tabular}
\label{tab:architecture}
\end{table*}

\begin{figure}[t]
\centerline{\includegraphics[width=\columnwidth]{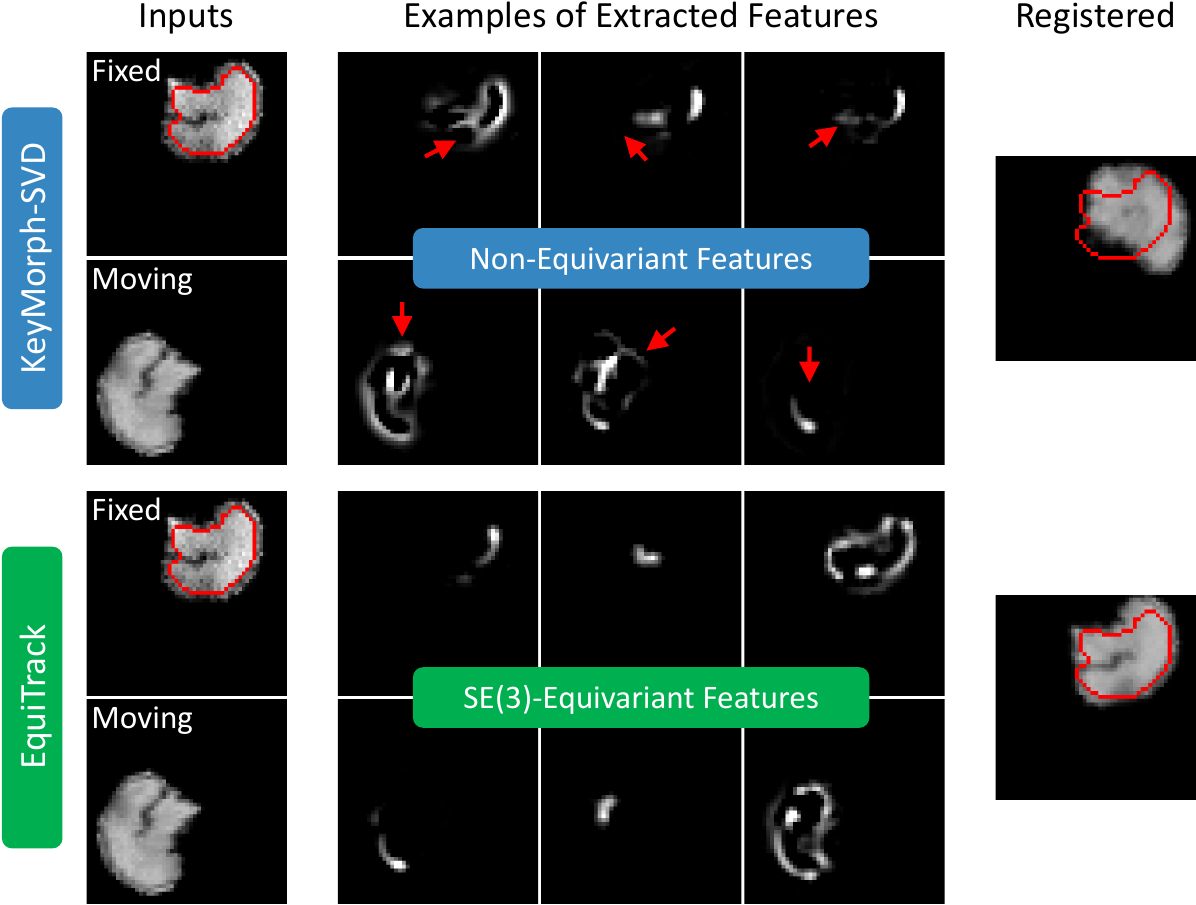}}
\caption{Example of features produced by KeyMorph-SVD and EquiTrack on a simulated pair from Fetal-I. The regular CNN of KeyMorph-SVD can lead to inconsistent representations across poses (red arrows). In contrast, The E-CNN used in EquiTrack guarantees the extraction of SE(3)-equivariant features, which enable accurate rigid registration.}
\label{fig:feats}
\end{figure}

\begin{figure*}[!t]
\centerline{\includegraphics[width=\textwidth]{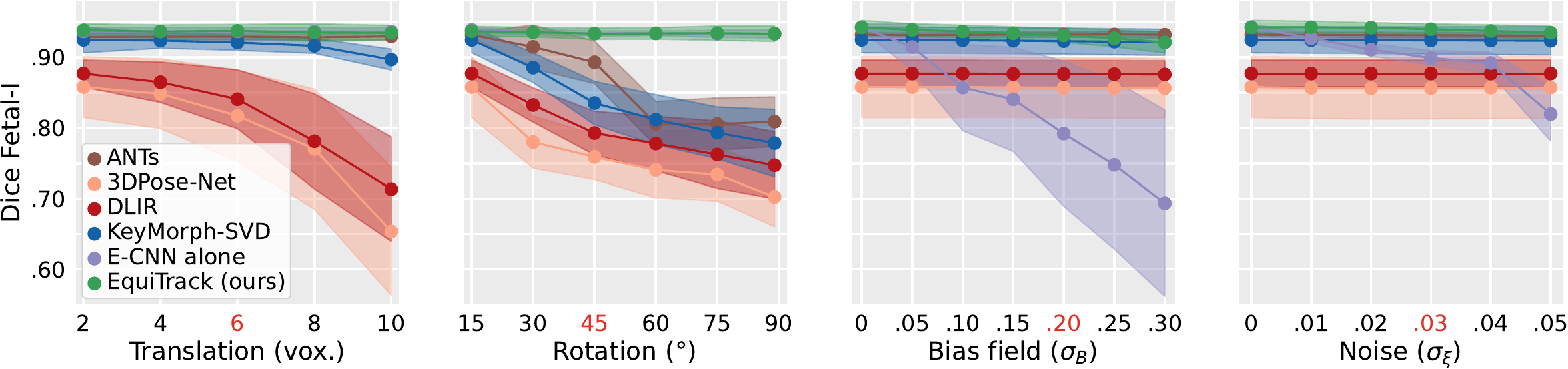}}
\caption{Sensitivity study (mean $\pm$ standard deviation) on Fetal-I for rotation, translation, bias field and noise. We start at translation $\!=\!$ 2 vox., rotation~$\!=\!$~15\dg, bias $\!=\!$ noise $\!=\!$ 0; and increase the effect of each augmentation separately. Red denotes the maximum values used to simulate~testing pairs in Fig.\ref{fig:boxplots}. E-CNN alone is a version of EquiTrack without a denoiser (the two methods have overlapping curves for the translation and rotation studies). In contrast to the baselines, EquiTrack is robust to both large motion (E-CNN) and intensity variations (denoising CNN) at the same time.}
\label{fig:sensitivity}
\end{figure*}

\section{Experiments and Results}
\label{sec:experiments}

\subsection{Testing on simulated pairs}
\label{sec:exp1}

The first experiment evaluates all methods on the adult and fetal MRI data. Here we use simulated pairs that provides us with \textit{exact} ground truth transforms. These are created from anchor volumes that we augment twice with more realistic intensity and spatial transforms than for training: rotations in [-45\dg, 45\dg], translations in [-6,6] vox., ${\sigma_B}_{\max} = 0.2$, $\sigma_\gamma = 0.2$, ${\sigma_\xi}_{\max}\! = \!0.03$ (these values are studied next~in a sensitivity analysis). In total, we generate 200 pairs from Adult and 300 pairs from Fetal-I. We measure accuracy by averaging translation and rotation errors along the three axes, and by computing Dice between the fixed and moved brain masks.

The results (Fig.\ref{fig:boxplots}) reveal that EquiTrack accurately tracks 3D rigid motion in adult and fetal brain MRIs, with mean angle and translation errors below 4\dg~and 0.5 voxels in all cases. Importantly, EquiTrack statistically outperforms all learning methods. While differences can be moderate on Adult (e.g., 2 Dice points with KeyMorph-SVD), the gap is considerably larger on Fetal-I with 10 Dice points. This emphasises the difficulty of analysing noisy fetal data, necessitating models with enhanced robustness. This is the case of our architecture, where using an E-CNN guarantees the extraction of matching features across poses (Fig.~\ref{fig:equivariance}). Indeed, in contrast with regular CNNs, the E-CNN used in EquiTrack enables consistent latent representations (Fig.~\ref{fig:feats}), which improve the quality of motion estimates compared to the other learning methods (Fig.~\ref{fig:example_exp1}). Meanwhile, we observe that processing images in parallel (i.e., DLIR vs. 3DPose-Net) is generally beneficial and is the solution implemented in EquiTrack. Interestingly, KeyMorph and KeyMorph-denoiser yield almost identical scores, showing that regular CNNs (as opposed to E-CNNs) do not need a denoiser to model noise invariance when trained with appropriate intensity augmentations. Finally, the fact that our closed-form algorithm accounts for potential landmark misalignments enables us to slightly improve the results of the original KeyMorph method: KeyMorph-SVD is better than KeyMorph by 1.7\dg~and 2.3\dg~on Adult and Fetal-I, respectively.

While the optimisation methods (RPM and ANTs) obtain results similar to EquiTrack on the Adult dataset (no statistical difference in rotation error and Dice), our method consistently outperforms them on Fetal-I by at least 2.5\dg~in rotation and 3.9 Dice points. We note that ANTs is slightly better than RPM for the fetal dataset, which highlights the benefits of exploiting intracranial intensity features in the images (ANTs), as opposed to purely relying on the segmentation mask (RPM). Importantly, EquiTrack is much more robust than ANTs on Fetal-I: the edge of rotation and translation error boxes extend to 30.6\dg~and 1.6 vox. for ANTs, as opposed to 4.6\dg~and 0.3 vox for EquiTrack. Also, our method is an order of magnitude faster than the optimisation methods (Table \ref{tab:architecture}), which is critical for online deployment of tracking methods.

\subsection{Sensitivity and ablation studies}
\label{sec:ablation}

In this section, we assess different aspects of our method, starting with its sensitivity to motion and noise. Specifically, we simulate testing pairs on Fetal-I by first using small augmentations (fixed rotation angles of 15\dg, shifts of 2 vox., $\sigma_B = \sigma_\xi = 0$) and increasing each value separately. Here we compare EquiTrack to the best variant of each baseline (ANTs for optimisation strategies, 3D-Pose-Net, DLIR, and KeyMorph-SVD), as well as an E-CNN alone (i.e., without a denoiser). Fig.\ref{fig:sensitivity} shows that the accuracy of learning methods quickly declines with larger motion, and especially rotations for which they lose at least 13 Dice points across the [15\dg, 90\dg] range. In contrast, the E-CNN alone maintains a very high accuracy across the studied ranges of translations and rotations, due to its SE(3)-equivariance. However, the E-CNN alone is not able to cope with increasing intensity augmentations (losing 24 and 13 Dice points in the bias field and noise sensitivity studies, respectively). On the other hand, the regular CNNs used in learning methods yield constant accuracy across the ranges of simulated bias fields and noises, which highlights their ability to learn noise invariance when trained with appropriate intensity augmentations, and confirms the results of the previous experiment (inefficiency of the denoiser in KeyMorph-denoiser). Crucially, EquiTrack is able to get the best of both worlds by combining the rigid equivariance of the E-CNN with the noise invariance of the denoising CNN, and yields scores above 94 Dice points in all cases.

\begin{table}[t]
\centering
\caption{Ablation study on Fetal-I with mean ($\pm$ standard dev.) scores.}
\setlength{\tabcolsep}{2.5pt}
\begin{tabular}{l c c c}
\hline
Method & Rot. err. (\dg) & Trans. err. (vox.) & Dice \\
\hline
E-CNN alone                                         & 22.4 (18.3) & 1.4 (1.3) & .84 (.15) \\
E-CNN (with intensity augm. $\zeta$)                & 20.2 (15.9) & 1.5 (1.5) & .86 (.12) \\ 
Denoiser (with $\zeta$) and E-CNN                   & 4.0 (3.5)   & 0.5 (0.3) & .93 (.04) \\ 
\hline
EquiTrack                                           & 3.9 (3.0)   & 0.4 (0.3) & .93 (.03) \\ 
\hline
EquiTrack, closed-form from \cite{wang_robust_2023} & 4.5 (6.1)   & 0.5 (0.8) & .92 (.11) \\
EquiTrack, no weights in SVD                        & 6.2 (4.2)   & 1.3 (0.9) & .90 (.06) \\ 
\hline
\end{tabular}
\label{tab:ablations}
\end{table}

\begin{figure}[t]
\centerline{\includegraphics[width=\columnwidth]{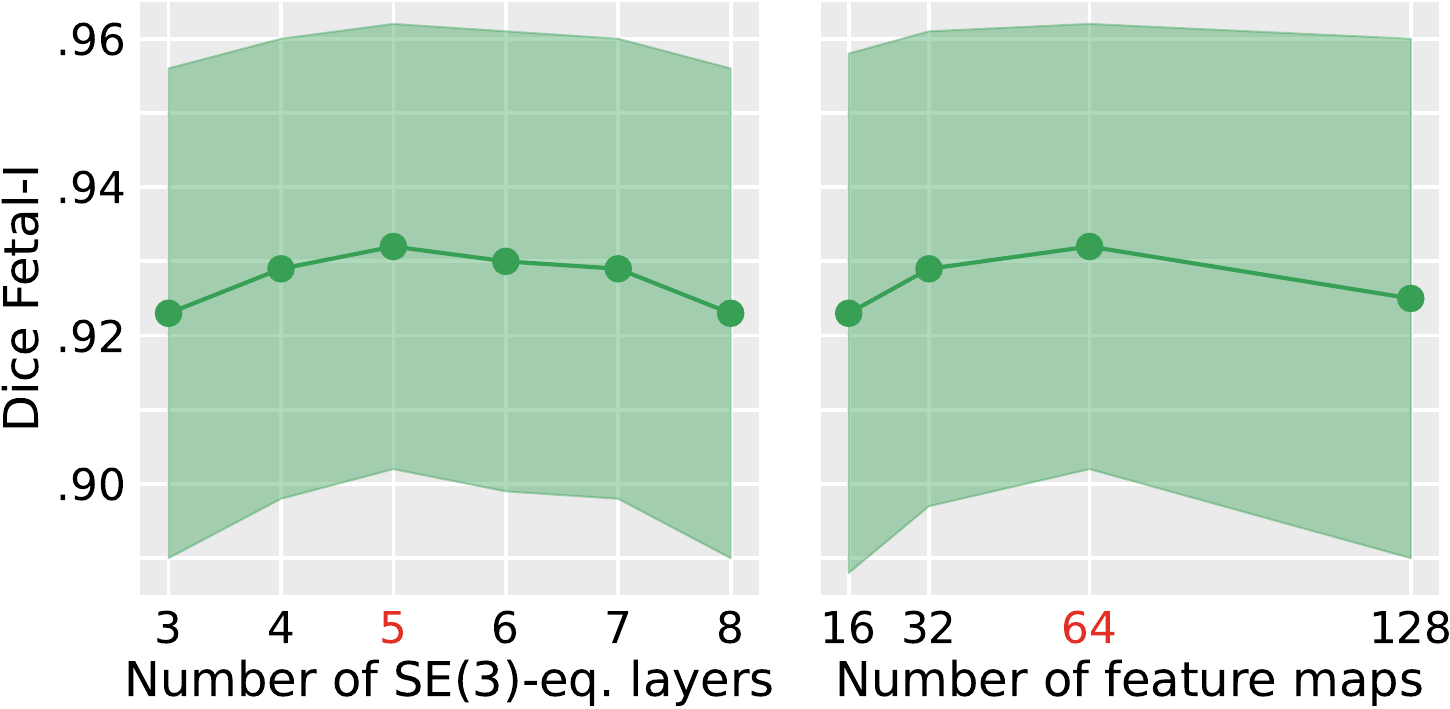}}
\caption{Sensitivity study on Fetal-I (mean $\pm$ standard deviation) for the number of SE(3)-equivariant layers and number $K$ of output feature maps of the pose regressor $\Phi$. Red indicates the default values, which were selected on an independent validation set (see Section~\ref{sec:architecture}).}
\label{fig:archi_ablations}
\end{figure}

We then conduct an architectural ablation study. Here, we start from the E-CNN $\Phi$ (i.e., the model used in our first work \cite{moyer_equivariant_2021}) and reconstruct EquiTrack by successively adding data augmentation (i.e., $\zeta$) for the inputs of $\Phi$, and the denoiser $\Psi$. Results are computed on the simulated fetal pairs of \ref{sec:exp1} (i.e., with intensity variations). Table \ref{tab:ablations} first expose that steerable E-CNNs alone cannot correctly register noisy fetal data (22.4\dg~angle error). Further, steerable E-CNNs have a limited learning capacity, since augmenting the inputs of $\Phi$ with the noise operator $\zeta$ only yields minor gains and remains highly inaccurate (20.2\dg~angle error). In contrast, adding the denoiser $\Psi$ (trained to undo $\zeta$ augmentations) largely enhances performance, as it provides the E-CNN with noise-free inputs that approximately satisfy $I^m \!= \!T \circ I^f$. Finally, we re-assemble all components, where using $\zeta$ to also train $\Phi$ leads to small but computationally cheap improvements (0.1\dg, 0.1~vox.).

We observe that replacing our closed-form algorithm with the KeyMorph variant yields more unstable results (+6 Dice points in standard deviation) as the latter does not account for potential landmark misalignments. Further, ablating the channel weights in our closed-form algorithm highlights the benefit of this mechanism (2.3\dg~in angle error).

Finally, we study the performance of EquiTrack as a function of the number of equivariant layers and output feature maps of $\Phi$. Figure~\ref{fig:archi_ablations} reveals that our method remains robust to architectural changes, with mean results varying by less than 0.8 Dice points across the ranges of studied values. The gradually decreasing scores when using more than 5 SE(3)-equivariant layers indicate that the limited learning capacity of E-CNNs is not a problem of number of learnable parameters.

\begin{table}
\centering
\caption{Mean ($\pm$ std.) statistics on real time series from Fetal-I and Fetal-II. KM is short for Keymorph-SVD. Best scores are in bold and marked with * if significant at 5\% (Bonferroni- corrected non-parametric Wilcoxon signed-rank test).}
\setlength{\tabcolsep}{2.2pt}
\begin{tabular}{l c c c c c c}
\hline
\multirow{2}{*}{Method} & \multicolumn{2}{c}{Rot. error (\dg)} & \multicolumn{2}{c}{Trans. error (vox.)} & \multicolumn{2}{c}{Dice}\\
 & Fetal-I & Fetal-II & Fetal-I & Fetal-II & Fetal-I & Fetal-II \\
\hline
ANTs      &          5.4 (4.2) &        7.0    (5.9) & \textbf{0.6}* (0.6) & \textbf{1.1} (1.1) &         .91   (.08) &         .89  (.15)\\
DLIR      &          9.9 (9.2) &        14.2  (11.7) &         1.3   (1.4) & 1.7 (2.4)          &         .86   (.09) &         .78  (.21)\\
KM        &          6.4 (6.4) &        12.1  (11.1) &         0.8   (0.8) & 1.6 (1.9)          &         .90   (.06) &         .82  (.19)\\
EquiTrack & \textbf{5.1}* (4.0)& \textbf{6.7}* (5.9) &         0.8   (0.6) & \textbf{1.1} (1.1) & \textbf{.92}* (.04) & \textbf{.90}* (.1)\\
\hline
\end{tabular}
\label{tab:scores_real}
\end{table}

\begin{figure*}[!t]
\centerline{\includegraphics[width=\textwidth]{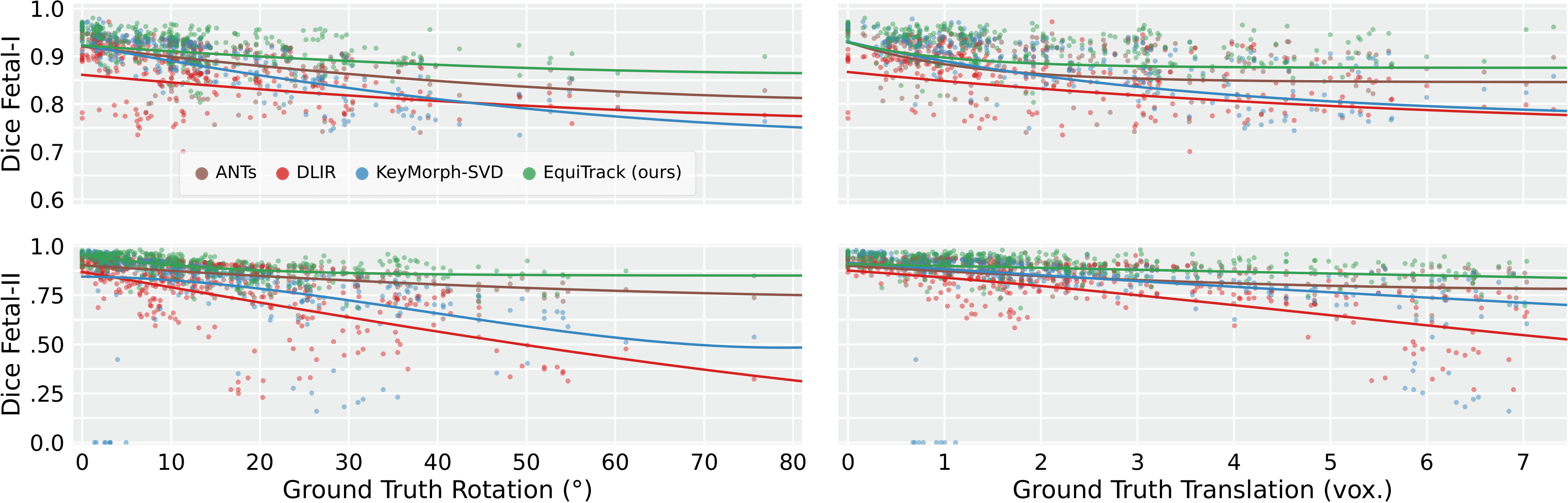}}
\caption{Dice scores as a function of ground truth rotations or translations for the Fetal-I and Fetal-II datasets. Trend lines are modelled with second order B-splines (6 equally spaced knots across value ranges), fit numerically with the L-BFGS-B method \cite{byrd_limited_1995} using the sum of squares of residuals.}
\label{fig:scatter}
\end{figure*}

\begin{figure*}[!t]
\centerline{\includegraphics[width=\textwidth]{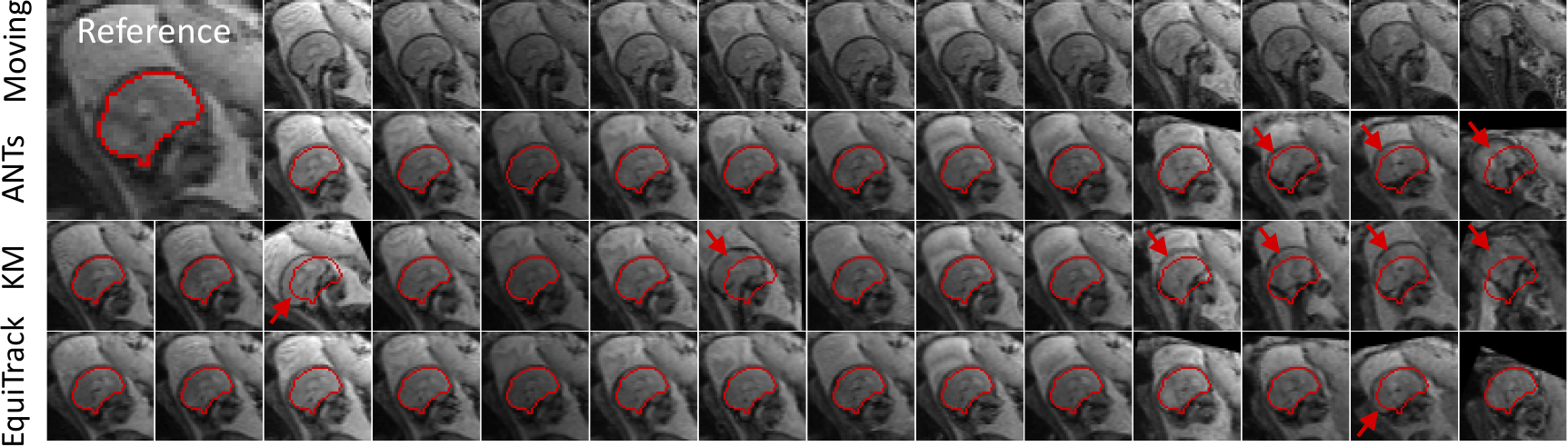}}
\caption{Illustration of brain motion tracking for a clinical time series from Fetal-II for ANTs, KeyMorph-SVD (KM) and EquiTrack. All time frames (top) are rigidly registered to the very first one (Reference). We overlay the reference brain mask on the results for visual assessment. Major mistakes are marked with arrows. While motion tracking was performed on \textit{masked} scans, we show the results on the \textit{unmasked} data to better visualise the scale of involved movements. Note how the fetus only starts substantially moving in the last four frames.}
\label{fig:example_exp2}
\end{figure*}

\subsection{Testing on real fetal time series}
\label{sec:exp_real}

The final experiment evaluates EquiTrack on \textit{real} time series from Fetal-I and Fetal-II, and compares it to ANTs, DLIR, and KeyMorph-SVD. For quantitative evaluation, we construct ground truth transforms by masking the original scans with segmentations of the brain and the eyes obtained with an in-house CNN, followed by ANTs registration to the first frame of the time series. We visually inspect the results and manually correct misregistrations with ITK-SNAP \cite{yushkevich_user-guided_2006}. Note that this process introduces a positive bias for ANTs.

Table \ref{tab:scores_real} presents the results, confirming the robustness of EquiTrack, whose scores are very similar to the previous experiments despite more complex contrast changes between frames. Further, EquiTrack only loses 2 Dice points when tested on the Fetal-II dataset (unseen during training), which highlights its generalisation across sequences (contrast and resolution) and populations (healthy vs. diseased). Importantly, EquiTrack significantly outperforms all baselines on both data- sets for all metrics (except ANTs for translations on Fetal-I).

Interestingly, the performance gap between EquiTrack and its competitors is smaller than in previous experiments. To better understand this result, we plot Dice scores as a function of motion (Fig.\ref{fig:scatter}), which reveals two findings. First, fetuses remain stationary most of the time, i.e., with motion below 15\dg/2 vox. for rotations/translations. In this case, all methods are fairly accurate. However, we observe an increasing gap between our method and other approaches as motion magnitude increases (Fig.\ref{fig:example_exp2}), confirming the results of the sensitivity analysis. For example, for angles above 30\dg, EquiTrack is better than KeyMorph-SVD by 8 Dice points on Fetal-I, and superior to ANTs by 7 Dice points on Fetal-II. Overall, EquiTrack surpasses all baselines, especially for larger movements, which are the main challenge in tracking fetal motion.

\section{Discussion}

We have presented a method for efficient tracking of 3D rigid motion. It leverages a steerable SE(3)-equivariant CNN to capitalise on natural symmetries arising in this task. We have seen in an ablation study that steerable E-CNNs alone are not expressive enough to learn noise invariance, even when trained with appropriate data (i.e., by applying data augmentation). As a result, when two time frames are affected differently by noise (which is common in medical imaging), an E-CNN might produce features that respond to different cues in the data, and thus are not equivariant anymore. We hypothesise that this limited capacity is due to the small number of learnable parameters, which only consist of the expansion coefficients for the precomputed basis filters. 

Since regular CNNs are much better at learning noise-robustness when trained with appropriate data, we introduce a denoising network to map noisy inputs to a common noise-free space. Hence, denoised volumes only differ by the pose of the brain, such that we can directly exploit the equivariance of the E-CNN to extract corresponding anatomical features. Our approach can be seen as a way to decouple the processing of intensity and anatomical features. These features can eventually be registered by computing summary statistics (i.e., centres of mass) that we align with a closed-form algorithm. Importantly, this algorithm is derived to minimise the overall distance between corresponding points, which makes it robust against slight misalignments that might have propagated through our framework. As a result, our method accurately tracks motion in challenging fetal MRI time series, even under substantial domain shifts in sequence, resolution, and population.

Overall, EquiTrack outperforms state-of-the-art optimisation and learning strategies for rigid motion tracking in brain MRI. Specifically, we show that these approaches might underperform when exposed to large motion, to which our method is robust by design. This outcome has already been reported for architectures similar to DLIR and 3DPose-Net, and to a lesser extent for optimisation strategies \cite{wang_robust_2023}. However, this is a more surprising observation for KeyMorph, which has obtained accurate results on adult brain MRI, even in the case of large misalignments \cite{wang_robust_2023}. This result suggests that the challenging fetal data (i.e., low tissue contrast, spherical shape with no easily identifiable geometry to guide the registration) requires highly robust architectures to be analysed, such as the hybrid two-step approach of EquiTrack.

While the proposed method yields state-of-the-art results in rigid motion tracking, it has limitations. First, the E-CNN expects the two images to be the same up to an unknown rigid transform. While we mitigate this constraint by using a denoiser for motion tracking, where we register the \textit{same} object seen under different poses, this is a much harder problem for inter-subject registration. Hence, our method has not been validated on general rigid registration. Further, by focusing on rigid motion, EquiTrack does not model non-linear local distortions caused by EPI sequences. However, this is not a problem in practice, since the scale of these distortions remains marginal compared to that of the rigid movements.

Another limitation is the lack of explainability of the E-CNN features. While KeyMorph was shown to identify keypoints corresponding to anatomical landmarks (i.e., points along region boundaries), our E-CNN extracts rather diffuse features that are hard to interpret. This observation lends more credence to the limited learning capacity of E-CNNs, which are not able to learn highly discriminative filters in our context.

Finally, we note the dependence of EquiTrack on the availability of a skull stripping method. However, we highlight that this dependence is not a problem specific to EquiTrack, since it is also a prerequisite for all baselines~\cite{mohseni_salehi_real-time_2019,wang_robust_2023,de_vos_deep_2019}. Moreover, skull stripping methods are readily available for adult brain MRI, and are already implemented in many fetal brain analysis pipelines~\cite{xu_fetal_2019,tourbier_efficient_2015,neves_silva_real-time_2023,mohseni_salehi_real-time_2019}.

\section{Conclusion}

We have presented a novel method based on equivariant networks for motion tracking. We demonstrated EquiTrack in the challenging scenario of fetal clinical time series, where it outperformed state-of-the-art learning methods and ANTs, while running an order of magnitude faster. The enhanced tracking accuracy and robustness were achieved by leveraging a two-step architecture to first remove anatomically irrelevant sources of noise in the inputs and then compute equivariant features for accurate motion prediction.

Future work will first focus on extending our approach to general rigid registration, possibly by combining standard and equivariant filters \cite{finzi_residual_2021} to simultaneously learn noise invariance and spatial equivariance. Further, while we focus on the fetal head, new \textit{piecewise}-rigid equivariant models may be beneficial to track the whole body \cite{chi_dynamic_2023}. More generally, future work will seek to tackle affine registration by incorporating scaling equivariance \cite{sosnovik_scale-equivariant_2020}. Preliminary methods have been proposed for scale-roto-translation equivariance \cite{gao_deformation_2022}, but making them tractable for large 3D volumes remains challenging. Another direction is to improve our representation of spatial features by diversifying the outputs of the E-CNNs with higher-order fields, which will need novel robust statistics. We will also seek to improve our augmentation model of EPI intensity variations by directly modelling them in closed form or by learning in a data-driven way. Finally, we will address the technical practicalities of deploying EquiTrack for slice prescription in 2D high-resolution fetal scans, where fast EPI volumes can be acquired to track inter-slices motion, possibly with automated detection of low quality scans \cite{scheinost_fetal_2018,gagoski_automated_2022}.

Overall, EquiTrack paves the way for further studies on equivariance in medical imaging and promises to enable applications for motion detection and prospective slice prescription.

\appendices
\renewcommand{\thetable}{A\arabic{table}}
\setcounter{table}{0}

\section{Evaluation of fetal skull stripping}
\label{sec:app_segm}

Skull stripping (i.e., brain segmentation) is a crucial preprocessing step for our pipeline and the baselines \cite{mohseni_salehi_real-time_2019,de_vos_deep_2019,wang_robust_2023}, because it enables the removal of the non-rigid organs surrounding the brain. Although automated skull stripping is commonly performed for the analysis of \textit{fetal} brain MRI~\cite{xu_fetal_2019,tourbier_efficient_2015,neves_silva_real-time_2023,mohseni_salehi_real-time_2019,khalili_automatic_2019}, existing methods have been less extensively evaluated than for the \textit{adult} brain. Therefore, this section assesses the automated fetal brain segmentation method used in this paper for preprocessing.

Here, fast automated fetal skull stripping is performed by training a segmentation UNet \cite{ronneberger_u-net_2015}. This network comprises 4 levels, each with 2 convolution layers of $3\!\times\!3\!\times\!3$ kernels, ReLU activations, and batch normalisation \cite{ioffe_batch_2015}. The first layer includes 16 feature maps, this number is doubled after each max-pooling, and halved after each upsampling. Predictions are obtained from a final convolution layer with a sigmoid activation. This network is trained with the same intensity and spatial augmentation as the denoiser $\Psi$, to which we add non-linear augmentation \cite{billot_synthseg_2023} and random flipping in all directions.

This segmentation network is trained, validated, and tested on a subset of images from Fetal-I for which we have manual ground truth brain segmentations. Importantly, we keep the same subject splits as in Section~\ref{sec:datasets}, which provides us with 74 (training), 13 (validation), and 39 (testing) images from 21, 5, and 12 time series, respectively. Moreover, we also include 73 testing images with ground truth from Fetal-II to evaluate the performance of the segmenter under domain shift.

\begin{table}[h]
\centering
\caption{Mean (std.) and worst-case Dice scores obtained by our skull stripping network for the preprocessing step.}
\setlength{\tabcolsep}{2.5pt}
\begin{tabular}{|l | c c | c c|}
\hline
 & \multicolumn{2}{|c|}{Fetal-I} & \multicolumn{2}{c|}{Fetal-II} \\
 & No dilation & 4-voxel dilation & No dilation & 4-voxel dilation \\
\hline
Mean Dice & 0.93 (0.02) & 0.94 (0.01) & 0.91 (0.02) & 0.93 (0.02) \\
Worst-case Dice & 0.88 & 0.91 & 0.86 & 0.89 \\
\hline
\end{tabular}
\label{tab:segm_scores}
\end{table}

Table~\ref{tab:segm_scores} shows that the segmenter produces high quality segmentations on Fetal-I with average and worst-case scores of 93 and 88 Dice points, respectively. Moreover, we see that our 4-voxel dilation strategy (which is applied to both the predicted and ground truth segmentations) further improves results (worst-case 91 Dice points) by compensating for small under-segmented areas in the original predictions. Finally, the segmenter is also robust against slight domain gaps, since it yields accurate scores on Fetal-II (worst-case 89 Dice points) when faced with intensity and morphological shifts.


\end{document}